\documentclass[twocolumn,secnumarabic,amssymb, nobibnotes, aps, prd,superscriptaddress]{revtex4-2}

\usepackage{caption}
\usepackage{graphics}
\usepackage{graphicx}
\usepackage{epstopdf}
\usepackage{subfigure}
\usepackage[ruled,linesnumbered]{algorithm2e}
\usepackage{appendix}
\usepackage{mathrsfs}
\usepackage{caption}
\usepackage{tabularx}
\usepackage{makecell}
\usepackage{amsmath}
\usepackage{hyperref}

\hypersetup{hidelinks,
colorlinks=true,
allcolors=black,
pdfstartview=Fit,
breaklinks=true}

\usepackage{multirow}
\usepackage{algpseudocode}
 \pdfoutput=1
 \UseRawInputEncoding
\begin{document}

\title{ Efficient Arbitrated Quantum Digital Signature with Multi-Receiver Verification}%

\author{Siyu Xiong}
\affiliation{
	School of Mathematical Sciences, Sichuan Normal University, Chengdu 610068, China}
	\affiliation{College of Advanced Interdisciplinary Studies, National University of Defense Technology,
		Changsha 410073, China}
\author{Bangying Tang}
\affiliation{College of Computer, National University of Defense Technology, Changsha 410073, China}
\affiliation{Strategic Assessments and Consultation Institute, Academy of Military Science, Beijing 100091, China}
\author{Hui Han}
\affiliation{College of Computer, National University of Defense Technology, Changsha 410073, China}
\author{Jinquan Huang}
\affiliation{College of Advanced Interdisciplinary Studies, National University of Defense Technology,
	Changsha 410073, China}
\affiliation{School of Electronics and Communication Engineering, Sun Yat-sen University, Shenzhen 518107, China}
\author{Mingqiang Bai}
\affiliation{
	School of Mathematical Sciences, Sichuan Normal University, Chengdu 610068, China}
\affiliation{Institute of Intelligent Information and Quantum Information, Sichuan Normal University,
		Chengdu 610068, China}
\author{Fangzhao Li}
\affiliation{College of Advanced Interdisciplinary Studies, National University of Defense Technology,
		Changsha 410073, China}
\author{Wanrong Yu}
\affiliation{College of Computer, National University of Defense Technology, Changsha 410073, China}
\author{Zhiwen Mo}
\affiliation{
	School of Mathematical Sciences, Sichuan Normal University, Chengdu 610068, China}
\affiliation{Institute of Intelligent Information and Quantum Information, Sichuan Normal University,
Chengdu 610068, China}
\author{Bo Liu}
\email[E-mail: ]{liubo08@nudt.edu.cn}
\affiliation{College of Advanced Interdisciplinary Studies, National University of Defense Technology,
	Changsha 410073, China}

\date{December 2024}%

\begin{abstract}
Quantum digital signature is used to authenticate the identity of the signer with information theoretical security, while providing non-forgery and non-repudiation services.
In traditional multi-receiver quantum digital signature schemes without an arbitrater, the transferability of one-to-one signature is always required to achieve unforgeability, with complicated implementation and heavy key consumption.
In this article, we propose an arbitrated quantum digital signature scheme, in which the signature can be verified by multiple receivers simultaneously, and meanwhile, the transferability of the signature is still kept.
Our scheme can be simplified performed to various quantum secure networks, due to the proposed efficient signature calculation procedure with low secure key consumption and low computation complexity, by employing one-time universal hashing
algorithm and one-time pad encryption scheme.
The evaluation results show that our scheme uses at least two orders of magnitude less key than existing signature schemes with transferability when signing files of the same length with the same number of receivers and security parameter settings.

\keywords{quantum digital signature, arbitrated quantum signature, quantum network, CW-pumped quantum key distribution.}

\end{abstract}
\maketitle

\section{Introduction}
Digital signature is an important cryptographic algorithm to ensure the non-forgerability and non-repudiation of transmitted digital messages.
In general, classical digital signature schemes \cite{Rivest,Elgamal1985,johnson2001} are constructed based on complex mathematical problems, which face severe security challenges due to the rapid development of quantum algorithms and quantum computer implementation technologies \cite{shor1994, kaye2007,arute2019}.
Different from classical digital signatures, quantum digital signature (QDS) is based on the principle of quantum mechanics to achieve information-theory security, and use the unique properties of quantum states such as non-clonability and unobstructibility to provide higher security and unforgeability for digital information. With the continuous development of quantum communication and quantum computing technology, QDS has a wider application prospect and will play an important role in finance, cloud computing and other fields.

The first QDS scheme was proposed in 2001 which theoretically demonstrated the feasibility of digital signatures using the principles of quantum mechanics, however, it is hard to be implemented due to the use of long-time registers and nondestructive state comparison \cite{gottesman2001}.
In 2012, P. J. Clarke et al. used 50:50 beam splitters to perform quantum comparisons of coherent states and gave an experimental demonstration of QDS \cite{Clarke2012}.
Afterwards, various theoretical QDS schemes \cite{Amiri2016,Puthoor2016,Yang2017,Thornton2019,Qin2022} and experimental QDS schemes \cite{Collins2014,Yin2016,Yin2017,Zhang2018,An2019,Richter2021,Roehsner2021} have subsequently emerged to eliminate these unrealistic requirements from the first QDS scheme.
However, these schemes can only sign one-bit message in each round, and multi-bit message can only be signed one bit by bit.
In 2023, H. L. Yin et al. proposed an efficient three-party QDS scheme based on secret sharing, one-time pad, and one-time universal$_2$ hashing, which can sign multi-bit messages in each round, and analyzed the security of the scheme using perfect keys \cite{Yin2023perfect} and imperfect keys \cite{Yin2023imperfect}.
Recently, QDS has become increasingly practical and is being combined with different quantum technologies \cite{OE2024,Yine-com2024,PRApplied2024}.
All of the above schemes are only for three-party signature scenarios, that is, one signer and two receivers.

Recently, quantum networks has become a reality \cite{joshi2020trusted,chen2021integrated,zheng2023multichip}, and the transferability of three-party QDS schemes can be used in specific metropolitan area networks to expand the three-party signature scene into multiple parties.
In addition, there are many theoretical studies on QDS schemes for more receivers \cite{arrazola2015, cai2019, Tanwar2020}.
R. Amiri et al. \cite{cryptoeprint:2016/739} proposed the multi-receiver QDS scheme based on universal hashing which enjoys several favorable properties such as short secret key lengths, short signature length, and high efficiency compared to the previous schemes.
In 2022, E. O. Kiktenko et al. \cite{kiktenko2022} proposed the practical multi-receiver QDS scheme that guarantees the authenticity and transferability of arbitrary length messages in a QKD network. This theoretical work strongly promotes the implementation of multi-party QDS in real QKD networks.
In the same year, Y. Pelet et al. \cite{pelet2022unconditionally} achieved the first experimental demonstration of multi-receiver QDS scheme with transferability in the eight-node quantum network \cite{joshi2020trusted}. This is also the first demonstration of multi-receiver QDS scheme in a large network.

In transferable multi-receiver QDS schemes, multiple receivers need to verify signatures one by one, which is not an efficient signature method for multi-node quantum networks.

QDS with an arbitrator can rely on the fixed and trusted arbitrator to verify and confirm the integrity and authenticity of a message, and can resist forgery and denial attacks more effectively, so as to achieve higher security. It is suitable for transaction scenarios with higher security requirements.
Existing QDS schemes with a fixed  arbitrator \cite{zeng2002arbitrated,li2009arbitrated,gao2011cryptanalysis,feng2020arbitrated,lu2022verifiable}, which are used to sign messages composed of quantum states, have been studied since a very early time.
Up to now, whether quantum messages can be signed is still in the stage of theoretical controversy \cite{curty2008comment,zeng2008reply,choi2011security, cryptoeprint:2018/1164}.

Based on the idea of fixing one arbitrator, we propose the arbitrated one-to-many quantum digital signature (AQDS) scheme for classical messages which can complete the signature verification of multiple receivers simultaneously using one-time universal$_2$ hashing in \cite{Yin2023perfect} and one-time pad.
In this paper, we consider the scenario where the signer needs to sign the same message to multiple receivers.
By setting up an arbitrator, our proposed scheme can still effectively guarantee this transferability when applied to one-to-one forwarding verification.
This work is suitable for all types of existing quantum key distribution (QKD) protocols, and here we model the scheme for QKD model with continuous-wave pumped (CW-pumped) entangled-photon sources \cite{neumann2021model}.
Moreover, based on the keys of different link obtained in laboratory demonstration and intercity demonstration of the eight-node fully connected network \cite{joshi2020trusted}, we give theoretical signature rate of the proposed scheme on this network.
The evaluation results show that our scheme uses at least two orders of magnitude less key than existing signature schemes with transferability when signing files of the same length with the same number of receivers and security parameter settings.

The paper is structured as follows:
in the section II, we give the general characteristics of QDS schemes.
And in the section III, we give the detailed process of our AQDS scheme with multy-receiver verification and its security proof in the section IV.
In the section V, the performance of the scheme is analyzed, mainly evaluating the key consumption, time requried under the CW-pumped QKD model and its performance in the specific eight-user quantum network.
Meanwhile, we compare the proposed scheme with existing multi-receiver schemes in this section.
Finally, in the section VI, we give a conclusion of our paper.

\section{General requirements}
Similar to classical digital signature \cite{B. Schneier1994}, following the rules of QDS schemes given in Ref.\cite{zeng2002arbitrated}, we give the following characteristics for QDS schemes.

(1) \emph{No honest abort.} The process for signing a message can be completed successfully if all participants in the scheme are honest.

(2) \emph{No forgery.} Neither the receiver nor a possible attacker are able to change the signature or the attached message after completion. The signature may not be reproduced as well.

(3) \emph{No repudiation.} The signer may not successfully deny the signature and the signed message.

(4) \emph{Firm assignments.} Each message is assigned anew to a signature and may not be separated from it afterwards.

(5) \emph{Quantum nature.} The signature involves purely quantum-mechanical features without a classical analog and is therefore by nature nonreproducible and may not be denied or forged.

\section{Arbitrated Quantum Digital Signature with Multi-Receiver Verification}

\subsection{Assumptions}

\begin{figure*}[thpb]
	\centering
	\includegraphics[width=\linewidth]{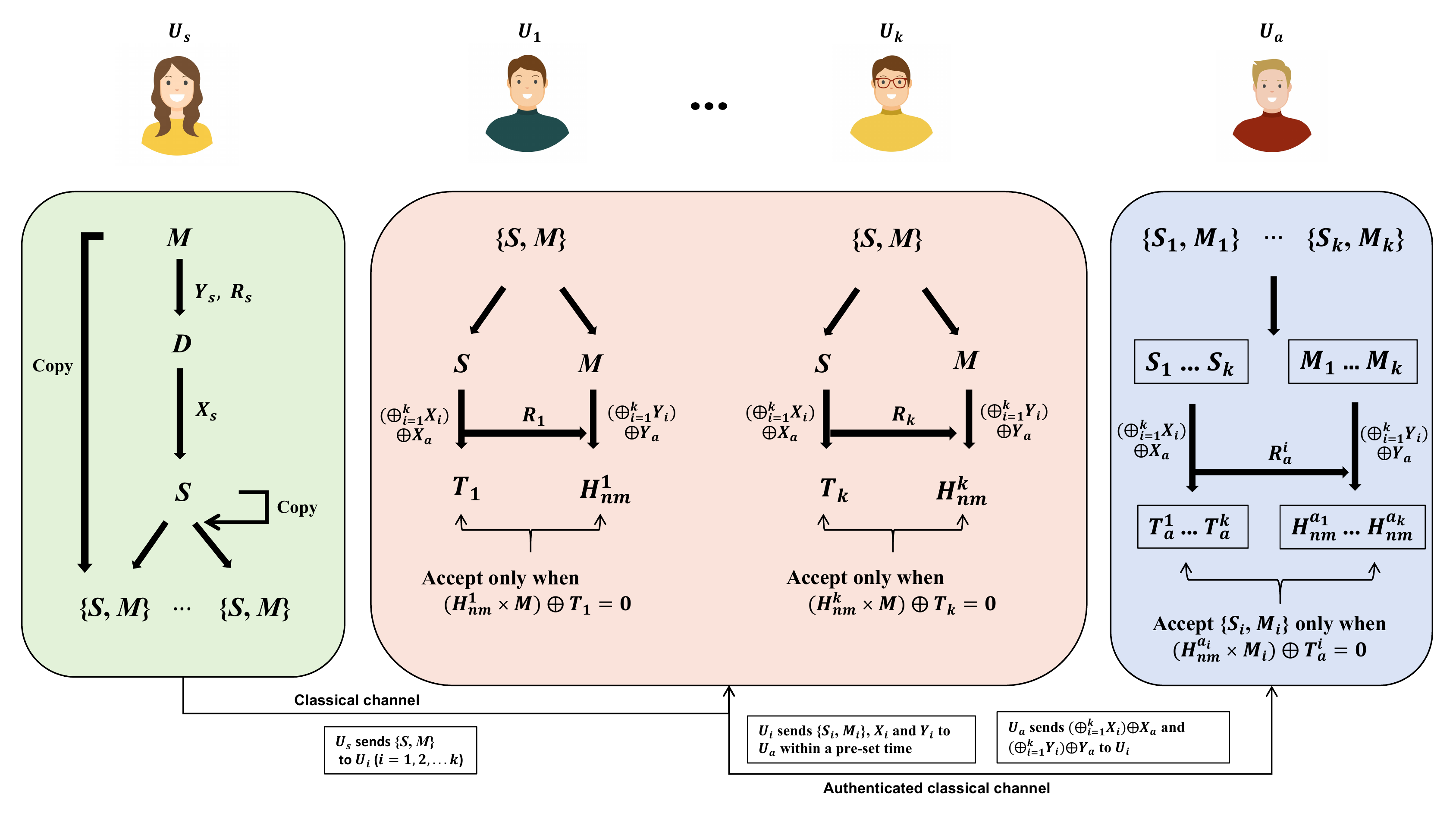}
	\caption{Schematic diagram of the AQDS scheme with multi-receiver verification. $M$ is the original message. $D$ is the digest generated by $U_s$, $S$ is the signature generated by the signer and $\{S_i, M_i\}$ ($1 \le i \le k,i \in N$) is the signature and message pair received by the arbitrator $U_a$ from  the receiver $U_i$.
		$X_i$ ($Y_i$) and $X_a$ ($Y_a$) are the $2n$ ($n$)-bit keys generated by QKD between $U_i$ and $U_s$ and between $U_a$ and $U_s$ respectively. $R_s$ is the $n$-bit quantum random number generated by $U_s$, $R_i$ is the random number obtained by $U_i$ decrypting $S_i$ from $U_s$, and $R_a^i$ is the random number obtained by $U_a$ decrypting $S_i$ from $U_i$.  $T_i$ ($T_{a}^{i}$) and $H_{nm}^k$ ($H_{nm}^{a_i}$) are the $n$-bit hash value and the Toeplitz universal hash function generated by $U_i$ ($U_a$), respectively. }
	\label{fig:aqds}
\end{figure*}

In this paper, the proposed arbitrated quantum digital signature (AQDS) with multi-receiver verification protocol is performed with the following assumptions.

\begin{itemize}
	\item The protocol is designed for the scenario that a signer $U_s$ signs a $m$-bit message $M$ , which will be sent to $k$ receivers $\{U_i\}_{i=1}^{k}$ respectively.
	\item The signature procedure requires the involvement of a constant trusted arbitrator $U_a$.
	\item All receivers will not actively interrupt the verification process.
	\item The quantum secure keys have to be prepared in advance between the signer and each receiver, as well as between the signer and the arbitrator.
	\item An authenticated classical channel has to be ensured between the signer and the arbitrator and between the arbitrator and each receiver.
\end{itemize}

The proposed AQDS scheme is mainly including three stages: the distribution stage, the messaging stage and the timeout verification and forwarding stage.

\subsection{Distribution stage}
In this stage, we assume that quantum key distribution protocol is perfectly executed between different users.

The signer $U_s$ and each user $U_i$, $i=1, 2, ..., k$ to generate a $3n$-bit secure key $K_{s,i}$, which can be described as
\begin{equation}
	K_{s,i} = (X_i,Y_i),
\end{equation}
where $X_i$ is performed for encrypting the generated digest and $Y_i$ is performed for generating the signature hash function. The length of secure key $X_i$ ($Y_i$) is $2n$ ($n$).

Meanwhile, a $3n$-bit secure key $K_{s,a}$ has to be generated between the signer $U_s$ and the trusted arbitrator $U_a$, which is given by
\begin{equation}
 	K_{s,a} = (X_a,Y_a),
\end{equation}
where the length of secure key $X_a$ ($Y_a$) is $2n$ ($n$).

Then, the signer $U_s$ calculates the encryption key $X_s$ and the signature key $Y_s$ by
\begin{equation}
	X_s = (\mathop  \oplus \limits_{i = 1}^k {X_i}) \oplus {X_a},
\end{equation}
and
\begin{equation}
	Y_s = (\mathop  \oplus \limits_{i = 1}^k {Y_i}) \oplus {Y_a}.
\end{equation}

\subsection{Messaging stage}

(a) \textit{Sign the message.}

The signer $U_s$ generates a string of $n$-bit quantum random numbers $R_s$, which is represented as an irreducible polynomial $P(x)$ of degree $n$. Take $Y_s$ as an initial vector, $U_s$ produces a Toeplitz universal hash function $H_{nm}$ with the linear feedback shift register (LFSR) structure. Then, the $n$-bit hash value $T$ of the message $M$ can be calculated by
\begin{equation}
T=H_{nm}\times M.
\end{equation}

Then, a $2n$-bit digest can be constructed by
\begin{equation}
D=(T||R_s).
\end{equation}

Afterwards, the signature $S$ can be calculated by
\begin{equation}
	S=X_s \oplus D.
\end{equation}

Finally, the message $M$ and the signature $S$ will be transmitted to $k$ receivers via the classical communication channel.

(b) \textit{Verification by the receiver.}

The receiver $U_i$ should obtain and then forward $M$, $S$ and the secure key $K_{s,i}$ to the arbitrator $U_a$ via the authenticated channel within a certain pre-set time range, where $1\leq i\leq k$. Assume the message and signature gained by $U_a$ are $\{M_i, S_i\}$, which may be forged by the receiver. Once the pre-set receiving time is over, $U_a$ will actively send the list of timeout receivers to the signer and get the keys of the timeout receivers from $U_s$ via the authenticated channel.

Then $U_a$ calculates the verification secure keys by
\begin{equation}
	{X_a^{'}} = (\mathop  \oplus \limits_{i = 1}^k {X_i}) \oplus {X_a},
\end{equation}
and
\begin{equation}
	{Y_a^{'}} = (\mathop  \oplus \limits_{i = 1}^k {Y_i}) \oplus {Y_a}.
\end{equation}

Afterwards, $X_a^{'}$ and $Y_a^{'}$ will be transmitted from $U_a$ to the receivers which forward the required information on time via the authenticated channel.

The receiver $U_i$ can decrypt an $n$-bit expected hash value and an $n$-bit random number, which can be expressed as
\begin{equation}
\left( {{{T_i}}||{R_i}} \right) =  X_a^{'} \oplus S.
\end{equation}
Take $Y_{a}^{'}$ as an initial vector, $U_i$ produces a Toeplitz universal hash function $H_{nm}^{i}$ with the LFSR structure.
If the result is
\begin{equation}
	\left( {H_{nm}^i \times M} \right) \oplus {{T_i}} = 0,
\end{equation}
the reciever $U_i$ announces that the signature from $U_s$ is accepted, if not, the signature is rejected.

(c) \textit{Verification by the arbitrator.}

If the reciever $U_i$ announces that the signature is accepted, the arbitrator $U_a$ get an $n$-bit  hash value and an $n$-bit random number by the decrypting operation
\begin{equation}
\left( {{{T_a^{i}}||{R_a^{i}}}} \right) = X_a^{'} \oplus S_i.
\end{equation}

Then $U_a$ uses $Y_a^{'}$ and $R_a^i$ to generate a Toeplitz universal hash function $H_{nm}^{a_i}$ with the LFSR structure.
If the result is
\begin{equation}
	\left( {H_{nm}^{a_i} \times M_i} \right) \oplus {T_a^{i}} = 0,
\end{equation}
the arbitrator $U_a$ announces that the signature to $U_i$ is successful; otherwise, the signature is invalid.

The process of the messaging stage is shown in FIG.~\ref{fig:aqds}.

\subsection{Timeout verification and forwarding stage}
After one time of the multi-receiver signature, the arbitrator has the signed message $M$, the signature $S$, and the key set $\left\{ {{K_{s,i}}} \right\}_{i = 1}^k$.

Assuming the $U_i$ is the timeout receiver, $U_i$ can forward the signature and message $\{S,M\}$ and the keys $\{X_i,Y_i\}$ to the arbitrator via the authenticated channel after the multi-receiver signature of the current round is completed.
The arbitrator can give a verification result directly based on whether the signature and message are the same as the signature and message in the completed round, and whether the key belongs to the key set.
The timeout receiver can decide whether to receive the signature directly based on the verification result of the arbitrator.

If $U_i$ is the receiver that has accepted the signature, $U_i$ can forward the signature and message to other users via the authenticated channel, and the arbitrator can also give a verification result.

\section{Security proof}

In this article, the hosest abort, forgery attack of the receivers and the repudiation attack of the signer are mainly considered for the security proof of the proposed AQDS scheme.
At the same time, the security analysis does not consider the attack by the possible attacker on the QKD protocol, that is, the attack occurs after the distribution stage.

\subsection{Robustness}

The robustness of the AQDS scheme quantifies the probability that at least one reciever rejects the signature when all participants are honest. Assume that all participants are honest, the key strings obtained after the XOR operation by the arbitrator and the key strings of the signer satisfies
\begin{equation}
	X_a^{'}=X_s, Y_a^{'}=Y_s.
\end{equation}

The random numbers representing irreducible polynomials decrypted by $k$ receivers and the arbitrator using the same key $Y_s$ satisfies

\begin{equation}
	{R_s} = R_i = R_a^i,
\end{equation}
where $1 \leq i \leq k$. All receivers and the arbitrator perform the identical universal$_2$ hash function to calculate the correct digest.

Therefore, the signature of the signer will be successfully received by $k$ recievers, and the probability of honest aborting is zero. Here we ignore the insignificant failure probability of classical bit error correction of quantum communication protocols in the distribution stage.

\subsection{Forgery}
Assume that the sender and the arbitrator are honest participants in the forgery attack of the receivers. The difference between individual forgery attack and joint forgery attack in the AQDS scheme is only the number of keys from the set $\left\{ {{K_{s,i}}} \right\}_{i = 1}^k$ held by the dishonest receivers.

The security of the key $X_s$ is guaranteed by the key $X_a$, thus the leaked information about $X_a$ to dishonest receivers will not be increased while the more keys in the set $\left\{ {{K_{s,i}}} \right\}_{i = 1}^k$ are leaked.
Therefore, the probability of a successful joint forgery attack is the same as the probability of a successful individual forgery attack.

Here we consider the scenario where the receiver $U_i$, in collusion with all other receivers, has obtained the key set $\left\{ {{K_{s,i}}} \right\}_{i = 1}^k$ and is ready to carry out a forgery attack.

The receiver $U_i$ can accomplish this attack by generating a faked message and signature that the signer $U_s$ has not signed, or by tampering with the message and signature from $U_s$.  If the arbitrator $U_a$ announces that the signature of $U_s$ forged by $U_{i}$ is valid, the forgery attack by $U_i$ is considered successful.

\textit{Case I}, the receiver $U_i$ has no information from $U_s$.

The only thing $U_i$ can do is guessing the key strings $X_s$, $Y_s$, $X_a$ and $Y_a$, then the successful probability of forgery attack can be calculated by
\begin{equation}
	\epsilon_{1} \leq \frac{1}{2^{n}}.
\end{equation}

\textit{Case II}, the reciever $U_i$ obtains the signature and message $\{S,M\}$ signed by $U_s$.

In this case, the reciever $U_i$ can execute an optimal attack by guessing out the random number string $R_s$. By randomly choosing an irreducible polynomial, the successful probability of obtaining the identical hash tag with two different messages by the reciever $U_i$ is given by \cite{Yin2023perfect}
\begin{equation}
	\epsilon_{2} \leq \frac{m}{{{2^{n - 1}}}}.
\end{equation}

Therefore, the successful probability of a forgery attack is
\begin{equation}
	{\epsilon _{f}} = \max \{ {\epsilon _1},{\epsilon _2}\}  \leq \frac{m}{{{2^{n - 1}}}}.
	\label{equ:ef}
\end{equation}

It should be noted that $U_i$ does not get any information from $U_a$ until $U_i$ forwards the message and the signature to $U_a$. Once $U_i$ chooses malicious delayed forwarding, $U_i$ will fail the authentication qualification of the current round due to the timeout issue.

\subsection{Repudiation}
A repudiation attack refers to a signer who signs a message and denies the signature in which at least one receiver and the arbitrator are honest participants.
When all receivers agree with repudiation of the signer, the signature is naturally invalid. Since the signer $U_s$ sends the signature of the message to all receivers, the signer cannot repudiate as long as one receiver validates a signed message sent by the signer. $U_s$ can successfully repudiate only if $U_a$ fails to verify the signature to all receivers.

Here, we consider the worst-case scenario where only one honest receiver $U_i$ is existed. Due to the procedure of the secure key exchange of the receivers and the arbitrator is protected by the authenticated channel, the arbitrator will always acquire the successful verification of the signature to the only one hosest receiver.
Therefore, the successful probability of the repudiation attack is $0$. Here, we ignore the insignificant failure probability of secure message authentication.

\section{Performance analysis}

The performance analysis of the proposed AQDS scheme is given on the assumption that all $k$ receivers forward the signature, the message, and keys to the arbitrator on time, and finally receive signatures within the valid time.

In this article, the secure key consumption, time required for multi-receiver signature, the signature performance in a specific eight-user metropolitan area network is analyzed for the proposed AQDS scheme.

\subsection{Secure key consumption}

In the AQDS scheme, $3n$-bit keys are required for each QKD link.
According to the Eq.~(\ref{equ:ef}), the parameter $n$ should at least satisfy
\begin{equation}
	n  \le  \log \frac{m}{\epsilon_f} + 1.
\end{equation}

The total number of keys required on all links can be expressed as
\begin{equation}
	{l} = 3n(k + 1) \le 3(\log \frac{m}{\epsilon_f} + 1)(k + 1).
\end{equation}

\begin{figure}[ht]
	\centering
	\includegraphics[width=\linewidth]{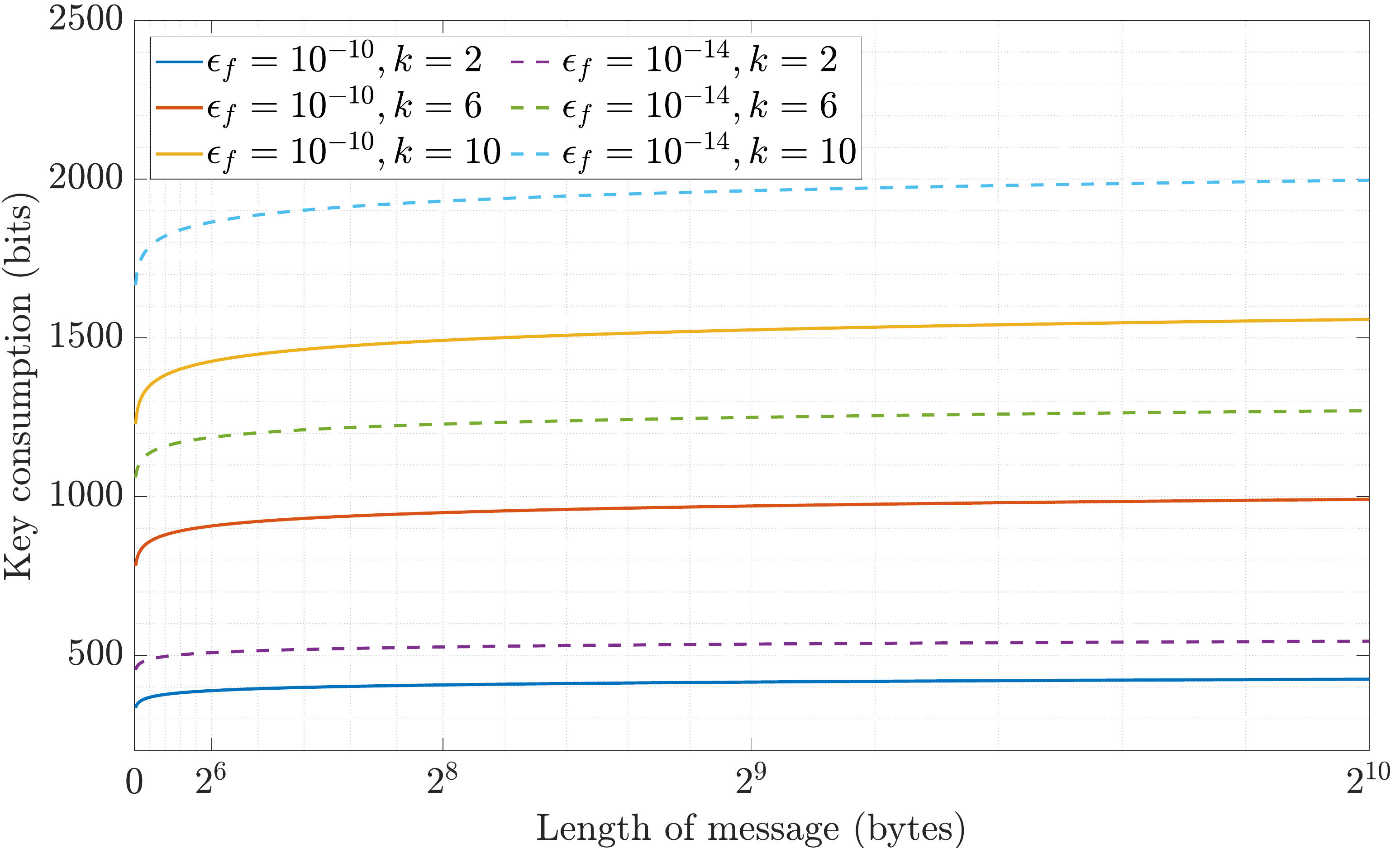}
	\vspace{1mm}
	\caption{{The secure key consumption versus the length of message. $\epsilon_f$ is the security parameter and $k$ is the number of receivers.}}
	\label{fig:consumption}
\end{figure}

When the security parameter $\epsilon_f$ is set as $10^{-10}$ and $10^{-14}$, and the number of recipients is $2$, $6$, and $10$, the total key consumption versus the length of message in shown the FIG.~\ref{fig:consumption}.
As shown in FIG.~\ref{fig:consumption}, the total secure key consumption is about $2000$ bits when ten receivers are signed with $\epsilon_f=10^{-14}$ and $m=2^{10}$ bytes.

\begin{figure}[ht]
	\centering
	\includegraphics[width=\linewidth]{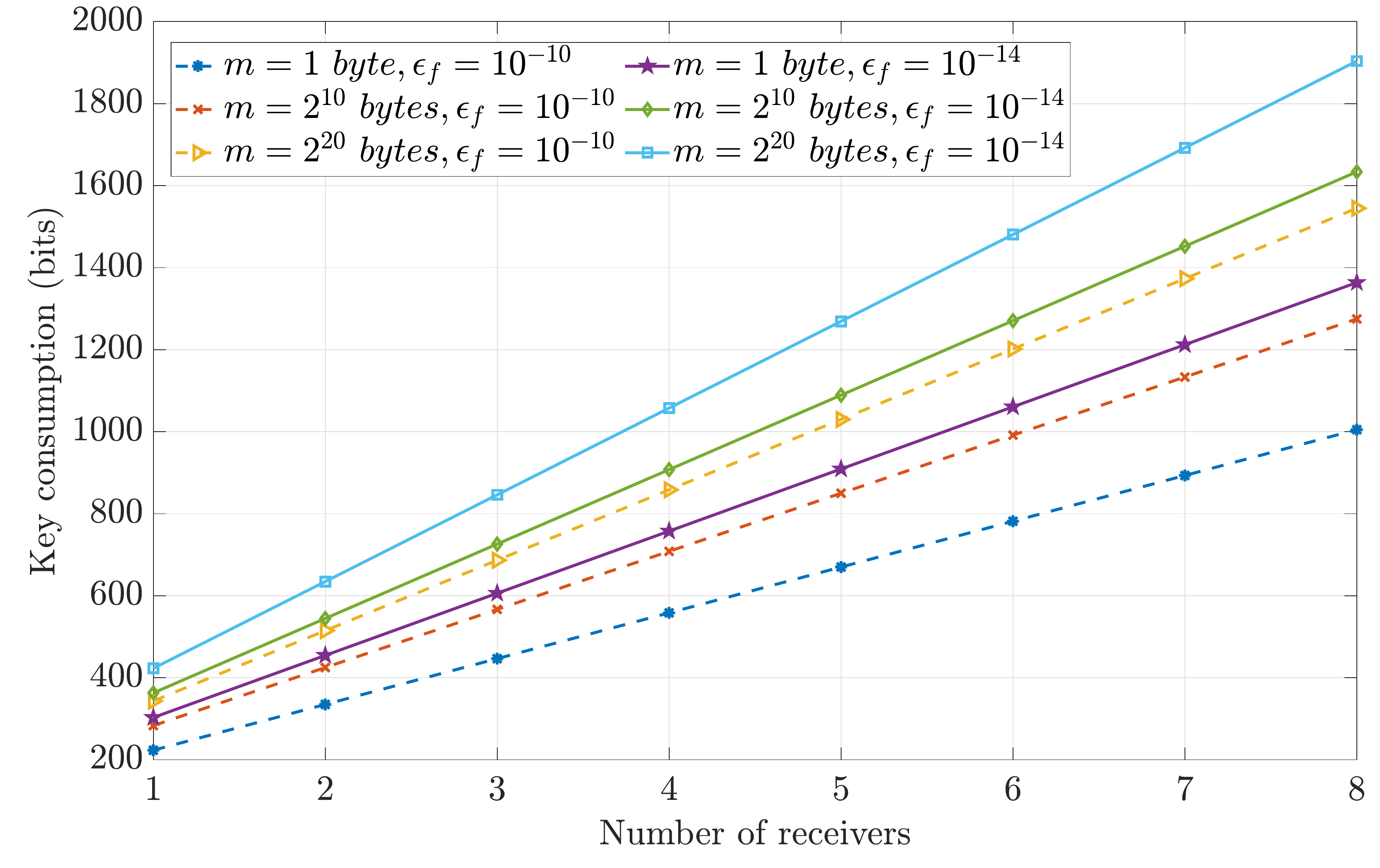}
	\vspace{1mm}
	\caption{{ The total key consumption versus the number of receivers. $\epsilon_f$ is the security parameter and $m$ is the length of message.}}
	\label{fig:amounts}
\end{figure}

When the security parameter $\epsilon_f$ is fixed as $10^{-10}$ and $10^{-14}$ respectively, the amount of keys required to sign $1$-byte, $1$KB, and $1$MB messages to different number of receivers is shown in FIG.~\ref{fig:amounts}.
As an example, the total consumption of keys is less than $2$K bits when $k=8$, $\epsilon_f=10^{-14}$ and $m=2^{20}$ bytes.

\subsection{Time requried for multi-receiver signature}
The QKD links in the proposed protocol can use all existing key generation protocols (KGP) to generate the required secure keys, such as BB84-KGP\cite{1984Quantum}, BBM92-KGP\cite{bennett1992phys} and TF-KGP\cite{lucamarini2018overcoming}.

\begin{figure}[ht]
	\centering
	\includegraphics[width=\linewidth]{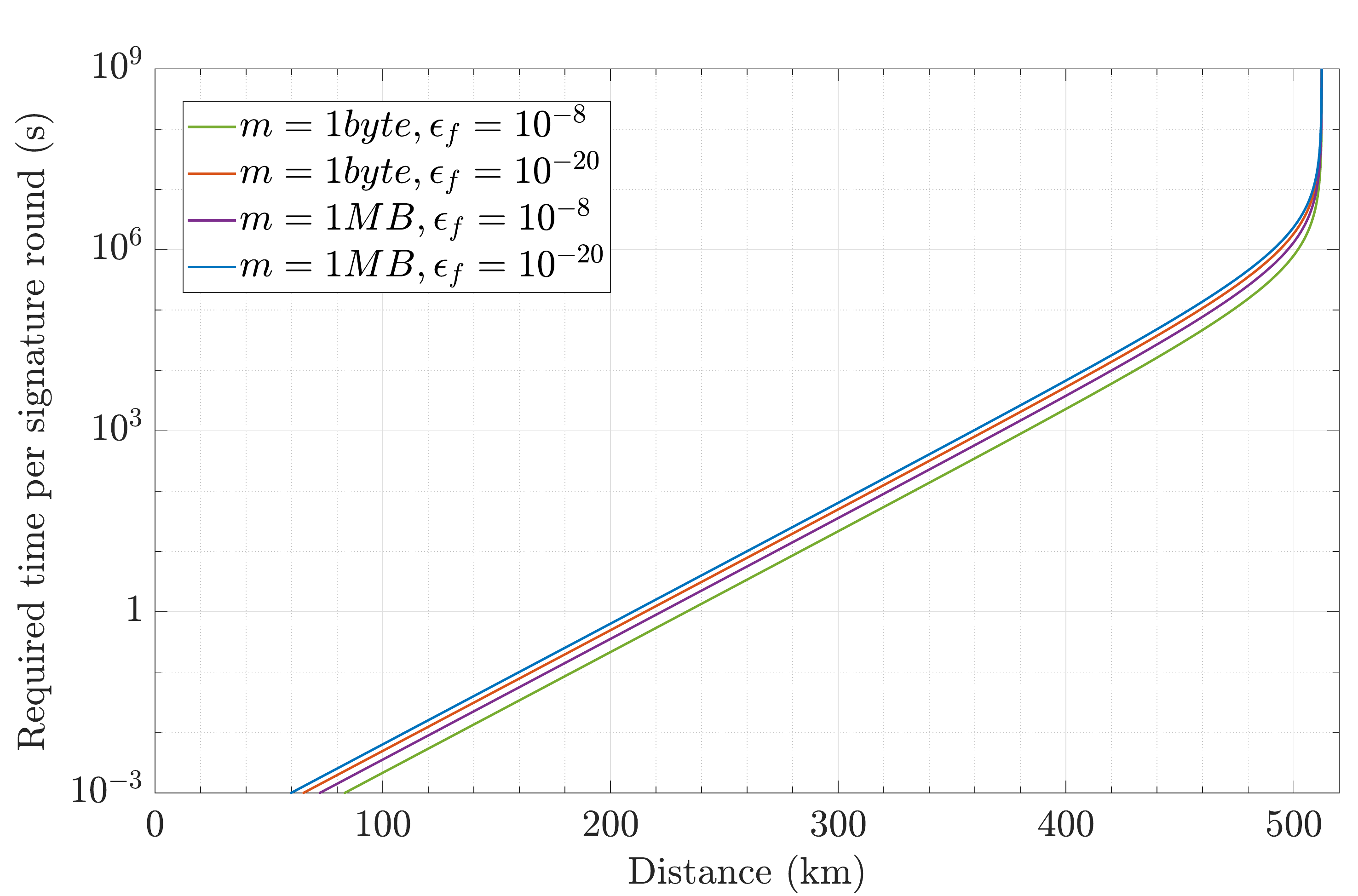}
	\vspace{1mm}
	\caption{{The time required to perform one round AQDS scheme with the message length of $1$ byte and $1$MB. }}
	\label{fig:time}
\end{figure}

Here we take BBM92 protocol as an example for modeling the simulation procedure, where  the continuous-wave pumped entangled-photon QKD \cite{neumann2021model} (CW-QKD) is proformed on each link. The calculation details of CW-QKD are in Appendix C.
In the simulation, the length of message is set to $1$ byte and $1$MB, respectively. The main simulation parameters are listed in Table~\ref{tab:para}.
The waiting time requried for performing the AQDS scheme is shown in FIG.~\ref{fig:time}.

\begin{table}[htbp]
	
	\caption{Simulation parameters.
	$ B $ is the brightness of the entanglement source.
	${{e^{pol}}}$ is the probability of individual polarization error.
	$DC$ indicates the dark count rate, and $t_{cc}$ indicates the width of window.
	$\eta$ is the loss at the receiving end.
	${{\eta ^{{t_{CC}}}}}$ is the proportion of true coincidences and
	$\alpha $ is the attenuation coefficient of fiber.}
    \label{tab:para}
	\begin{ruledtabular}
		\begin{tabular}{cccc}
			$B$&$e^{pol}$&$DC$& $\eta$ (dB) \\
			\hline
			 {\footnotesize $1 \times {10^8}$}&{\footnotesize 0.0181}&{\footnotesize 300 }& 3\\
			\hline\hline
			${\eta ^{{t_{CC}}}}$&$\alpha$ (dB/km)&$t_{CC}$ (ps)&\\
			\hline
			 {\footnotesize 0.761}&{\footnotesize 0.2 }&{\footnotesize 500 }&\\
		\end{tabular}
	\end{ruledtabular}
\end{table}

As shown in the figure, for example, when the link distance between the signer and other users is $360$ km, the security parameter is set to $10^{-20}$, and the time required for executing the AQDS scheme to sign a $1$MB message is about $10^{3}$ seconds.

\begin{figure}[htbp]
	\centering
	\includegraphics[width=\linewidth]{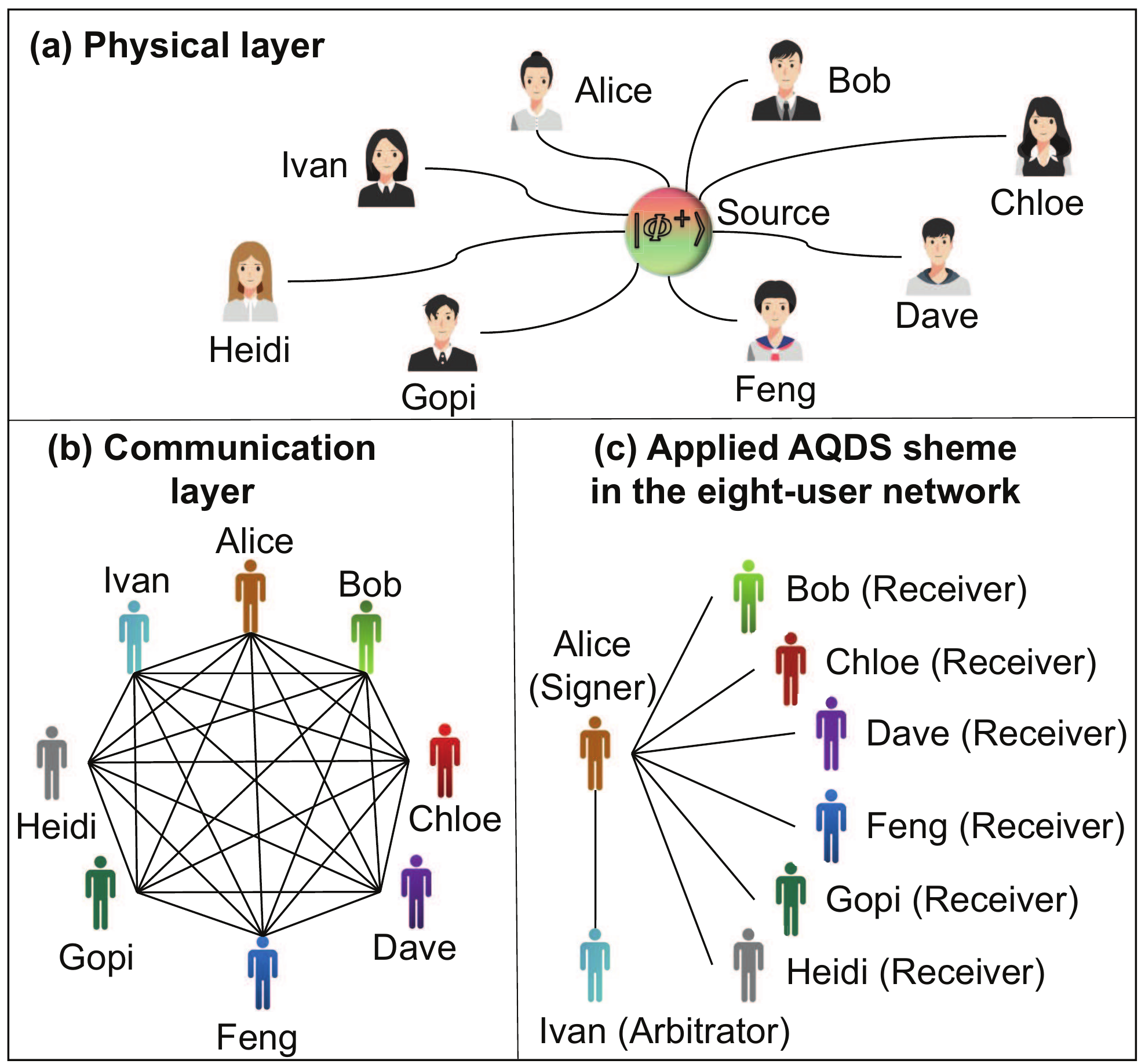}
	\vspace{1mm}
	\caption{{\small Multi-receiver AQDS in the eight-user quantum network. (a) The physical layer topology of the eight-user quantum network. (b) The communication layer topology of the eight-user quantum network. (c) AQDS scheme in the eight-user network. In this scenario, Alice is the signer, Ivan is the arbitrator, and the other six users are the receivers.}}
	\label{fig:eightusernetwork}
\end{figure}

\subsection{Performance in the specific  eight-user quantum network}

\begin{table*}[!htbp]
	\caption{Performance in the specific eight-user quantum network.}
\label{tab:per}
\begin{ruledtabular}
\begin{tabular}{c|c|c|c|c|c|c|c|c|c|c|c|c}
	\multirow{3}{*}{}     & \multicolumn{8}{c|}{The secure key distribtion stage}                       & \multicolumn{4}{c}{The performance of the AQDS scheme}                                                                                                                           \\
	\cline{2-13}
	Scenario & \multirow{2}{*}{\thead{Total\\ time}} & \multicolumn{7}{c|}{Secure key per link (bits)} & \multirow{2}{*}{\thead{$m$\\ (bytes)}} & \multirow{2}{*}{$\epsilon_{f}$} & \multirow{2}{*}{\thead{Secure keys used for \\signature per link} }&  \multirow{2}{*}{\thead{Supported rounds \\for signature} }\\
	\cline{3-9}
	&& AB & AC & AD & AF & AG & AH & AI &&&&\\
	\hline
	\thead{Laboratory \\demonstration}           &       {\footnotesize 18.45 h}      &  {\footnotesize 10.03M}  & {\footnotesize 10.08M}      &  {\footnotesize 9.58M}  &  {\footnotesize 13.37M}  &  {\footnotesize 16.53M}  &  {\footnotesize 9.06M}  &       {\footnotesize 6.81M}            &     \multirow{2}{*}{\thead{\\$2^{10}$}}                &  \multirow{2}{*}{\thead{\\$10^{-10}$}}                                               & {\footnotesize 6.81Mb} &         {\footnotesize49589}                                            \\
	\cline{1-9} \cline{12-13}
	\thead{Metropolitan \\network \\demonstration} &         {\footnotesize 27 min}      &  {\footnotesize 31143}  & {\footnotesize 8926}      &  {\footnotesize 6087}  &  {\footnotesize 15590}  &  {\footnotesize 38075}  &  {\footnotesize 6637}  &       {\footnotesize 901}                  &                                                  &                    &  {\footnotesize 901 bits}  &  {\footnotesize 6}
\end{tabular}
	\end{ruledtabular}
\end{table*}

\begin{table*}[!htbp]
	\caption{Comparison result of different multi-receiver QDS schemes.
		$k$ is the number of receivers and $\xi$ is the integer parameter which is a significant factor in determining the security level and transferability of the signature scheme.
		$m$ and $\epsilon_{f}$ are the message length and security parameters of the signature, respectively.
		$K_{total}$ is the total consumption of keys.}
\label{tab:com}
	\begin{ruledtabular}
		\begin{tabular}{ccccccc}
            \multirow{2}{*}{Scheme} & \multicolumn{2}{c}{Time complexity} & \multicolumn{4}{c}{Consumed security key}\\

            \cline{2-3}\cline{4-7}

			 & {Sign}  &Verification & $k$ &$m$ (bit) &$\epsilon_{f}$ & ${K_{total}}$ (kbit)\\

			\hline\hline

            {\footnotesize Amiri et al. \cite{cryptoeprint:2016/739}}&\multirow{3}{*}{{\footnotesize {$O\left( {{k^2}\xi *m} \right)$}}}&\multirow{3}{*}{{\footnotesize {$O\left( {k\xi *m} \right)$}}}&{\footnotesize {$7$}}&{\footnotesize {$8$}} &{\footnotesize {$10^{-10}$}}&{\footnotesize {$21.888$}}\\

            \cline{1-1}\cline{4-7}

			 {\footnotesize Pelet et al. \cite{pelet2022unconditionally}}&&&{\footnotesize $7$}&{\footnotesize $8$} &{\footnotesize $10^{-10}$}&{\footnotesize $35.898$}\\

			\cline{1-1}\cline{4-7}

			 {\footnotesize Kiktenko et al. \cite{kiktenko2022}}&&&{\footnotesize $4$}&{\footnotesize $8$M}
&{\footnotesize $10^{-10}$}&{\footnotesize $279.400$}\\

			\hline

				{\footnotesize Extended Yin et al. \cite{Yin2023perfect}} &\multirow{2}{*}{{\footnotesize {$O\left( {k*m} \right)$}}}&\multirow{2}{*}{{\footnotesize {$O\left( {m} \right)$}}}&{\footnotesize $7$}&{\footnotesize $8$}&{\footnotesize $10^{-10}$}&{\footnotesize $1.596$}\\

				{\footnotesize  to multiple receivers}&&&{\footnotesize $4$}&{\footnotesize $8$M }&{\footnotesize $10^{-10}$}&{\footnotesize $1.392$}\\

				\hline

				 {\footnotesize AQDS with} &\multirow{2}{*}{{\footnotesize {$O\left( {m} \right)$}}}&\multirow{2}{*}{{\footnotesize {$O\left( {m} \right)$}}}&{\footnotesize $7$}&{\footnotesize $8$} &{\footnotesize $10^{-10}$}&{\footnotesize $0.912$}\\

				{\footnotesize multi-receiver verification}& & &{\footnotesize $4$}&{\footnotesize$8$M} &{\footnotesize $10^{-10}$}&{\footnotesize $0.870$}\\
			\end{tabular}
		\end{ruledtabular}
	\end{table*}

Here, we show the performance of the proposed AQDS scheme in the demonstrated eight-user metropolitan quantum entanglmement distribution network, which is detailed described in Ref.~\cite{joshi2020trusted}. Each pair of the eight users shares a different biparitite entangled state, forming the fully connected network the AQDS shceme required.

The quantum network can be divided into a physical layer and a communication layer. In the pysical layer, the entangled photons are transmitted to each user with a fiber link, using $16$ ITU communication wavelength channels to fully interconnect eight users in total, shown in FIG.~\ref{fig:eightusernetwork}(a). In the communication layer, a fully connected entanglement distribution network is naturally formed among eight users, shown in FIg.~\ref{fig:eightusernetwork}(b). Here, we apply the AQDS scheme by designating Alice as the signer, Ivan as the arbitrator, and the rest users as the receiver, as shown in FIG.~\ref{fig:eightusernetwork}(c).

The evaluated performance in the specific eight-user quantum network is given in Table~\ref{tab:per}. In the laboratory demonstration case, the modified BBM92 protocol was performed between each user pair and gained at least 6.81Mb secure key between Alice and other users in 18.45 hours. Assume the message length is $1$KB and $\epsilon_f = 10^{-10}$, the AQDS scheme can be perfromed 49589 rounds in total. In the Metropolitan demonstration case, the minimum amount of generated secure key was 901 bits between Alice and Ivan via a 15.74km fiber link in 27 minutes. Assume the message length is $1$KB and $\epsilon_f = 10^{-10}$, the AQDS scheme can be perfromed 6 rounds in total.

\subsection{Comparison}

In multi-receiver QDS schemes without an arbitrator, achieving a higher level of transferability often means consuming more security keys (see Appendix A.1).
As described in section III.4, the key required of our AQDS scheme does not vary with the transferability of the signature.

As shown in Table \ref{tab:com}, the key consumption of our signature scheme can save at least two orders of magnitude compared with minimum key consumption (when the transferability level is 1) of multi-receiver signature schemes with transferability  \cite{cryptoeprint:2016/739, pelet2022unconditionally, kiktenko2022} under the same number of receivers, the same length of message, and the same security parameter settings.

Meanwhile, our scheme can save about one-third of the key usage compared to the extended signature scheme of Yin et al. \cite{Yin2023perfect} where one receiver is a fixed trusted third party (see Appendix A.2).

\section{Conclusion}
We propose the arbitrated quantum digital signature scheme in which multiple receivers can verify signatures from the same signer simultaneously.
The key consumption of our signature scheme can save at least two orders of magnitude compared with recent multi-receiver signature schemes with transferability under the same parameter settings.
Our scheme has the fixed arbitrator, which is similar to the certificate authority and can be applied to some specific application scenarios with a trusted third party.
By using our scheme, multiple receivers can achieve low-key consumption authentication of the same signature simultaneously, and the scheme can be practiced in networks using any type of QKD protocol.

Quantum digital signature is becoming more practical and gradually oriented towards large-scale quantum networks. In the future, practical application systems based on quantum digital signature can be designed and implemented to verify their performance and security in real environments, so as to improve the architecture of quantum cryptographic protocols. High-security quantum digital signature can be applied to the application scenarios of finance, government affairs, communication and other fields to meet potential requirements. At the same time, quantum digital signature can be combined with emerging technologies such as the Internet of Things and blockchain to expand its application range.
\appendix

\section{Multi-receiver signature}
\begin{figure*}[t]
	\centering
	\includegraphics[width=\linewidth]{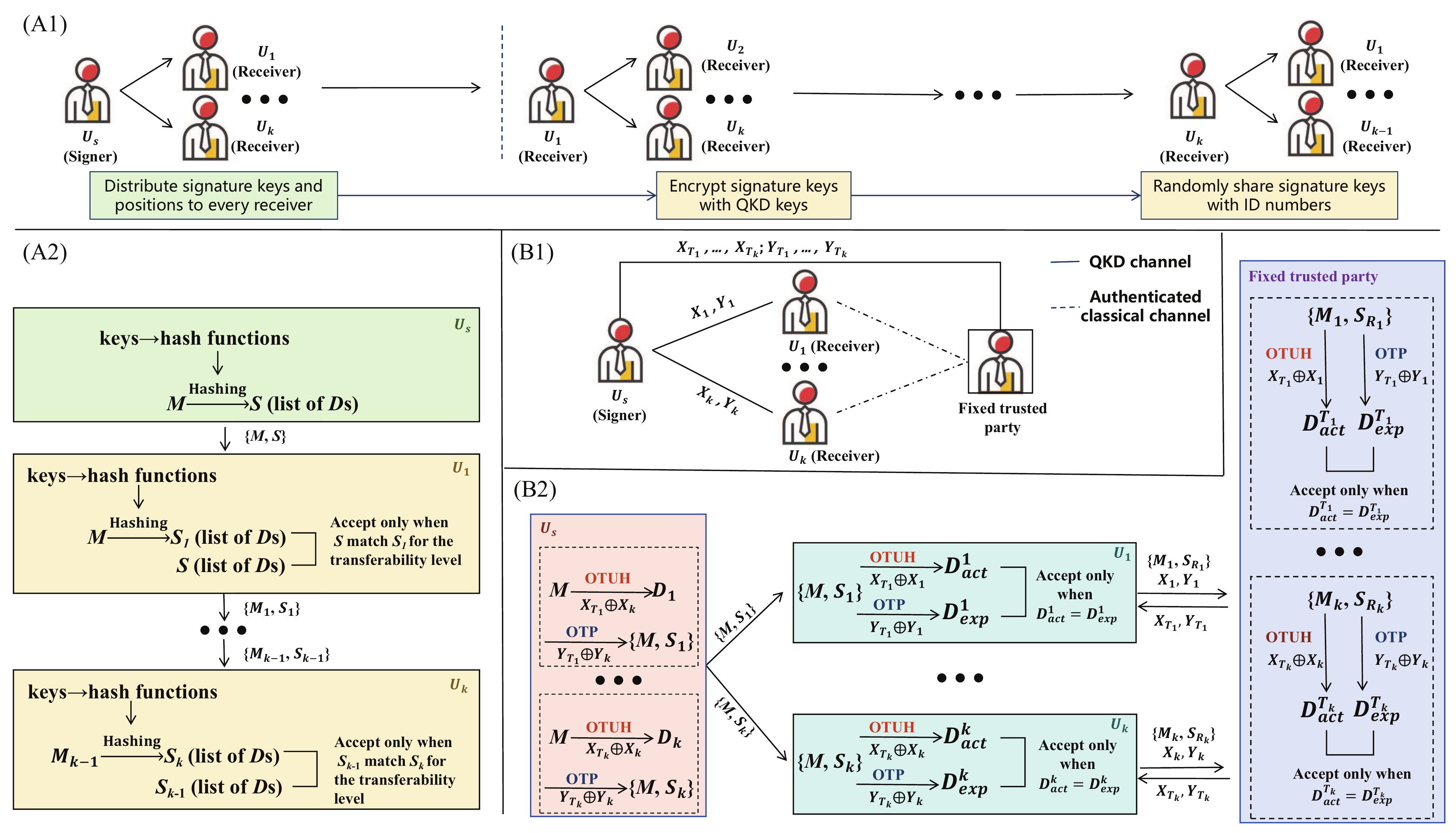}
	\vspace{1mm}
	\caption{\small Two types of QDS schemes for multiple receivers. (A1)  The distribution stage of multi-receiver QDS scheme without an arbitrator.
		(A2)  The messaging stage of multi-receiver QDS scheme without an arbitrator.
		$M$ is the original signed message and $S$ is the signature consisting of the list of the digests $D$s generated by the signer.
		$S_i$ ($i=1,...,k$) is respectively the signature consisting of the list of the digests $D_s$  generated by the receiver $i$.
		$\{M_i,S_i\}$ is the signature and message pair that the receiver $i$ forwards to the next receiver.
		(B1)  The distribution stage of extended QDS scheme with a fixed trusted party.
		$X_{T_1}$, ..., $X_{T_k}$($Y_{T_1}$, ..., $Y_{T_k}$) are $k$ $2n$ ($n$)-bit keys generated between the signer and the fixed trusted party through the QKD channel.
		$X_i$ ($Y_i$) is the $2n$ ($n$)-bit key generated between the signer and the receiver $i$ through the QKD channel.
		(B2)  The messaging stage of extended QDS scheme with a fixed trusted party for multiple receivers.
		$M$ is the signed message.
		$D_1$,..., $D_k$ and $S_1$,..., $S_k$ represents $k$ digests and $k$ signatures generated by the signer respectively.
		$D_{act}^i$ and $D_{exp}^i$ are the actual digest and the expected digest generated by the receiver $i$, respectively.
		$\{M_i, S_{R_i}\}$ is the message and signature pair sent by the receiver $i$ to the fixed trusted party.
		$D_{act}^{T_i}$ and $D_{exp}^{T_i}$ are the actual digest and the expected digest generated by the fixed trusted party using the message and signature pair $\{M_i, S_{R_i}\}$, respectively.
	}
	\label{fig:com}
\end{figure*}

\subsection{Multi-receiver quantum digital signature scheme with transferability}
Here, we take the scheme in the literature \cite{cryptoeprint:2016/739} as an example to give a brief introduction to this type of schemes.
The scheme is divided into two stages: the distribution stage (see FIG.~\ref{fig:com}(A1)) and the messaging stage (see FIG.~\ref{fig:com}(A2)).

\textbf{Distribution stage}
The signer $U_s$ generates a set of $k^2\xi $ keys as the signature keys, where $\xi$ is the parameter that determines the security level and transferability of the signature scheme.
This set of keys can be generated by the signer with $k$ receivers via QKD, each of whom has $k\xi$ keys.
Each receiver uses his encrypted channel (protected by QKD keys) to randomly exchange $\xi$ keys and their respective key identification (ID) numbers from his original set to each other.

\textbf{Messaging stage}
The signer $U_s$ generates multiple hash functions using the key generated in the distribution stage. These hash functions are applied to the message $M$ to generate the list of $k^2\xi$ digests (Ds). The list formed by Ds is the signature $S$, which is sent with $M$ to the receiver $U_1$. The receiver $U_1$ generates multiple hash functions using the key obtained during the distribution stage, and obtains the list of $k\xi$ Ds as the signature $S_1$ by hashing the message $M$ several times. Whether to accept the signature is determined by comparing the matching degree between $S$ and $S_1$ to meet the preset transferability level $l_{max}$.
(The matching degree of different transferability levels can be set with different parameters evenly spaced between $0$ and $0.5$.)
After that, the receiver $U_i$ ($2 \le i \le k - 1,i \in N$) can pass the signature to the receiver $U_{i+1}$, just as the  receiver $U_{i-1}$ passed it to him.
The receiver $U_{i+1}$ also determines whether to receive the signature by comparing the matching degree between the received $S_i$ and $S_{i+1}$ he generates.

\subsection{Extended three-party quantum digital signature scheme with a fixed trusted party}

In the security assurance of a three-party QDS schemes, it is often necessary to assume that at least two participants are honest (that is, at least one receiver is honest), although the honest person is not specified.
Here we take the scheme in \cite{Yin2023perfect} as the example to give a method to extend the three-party signature scheme to the multi-party signature scheme after fixing one receiver as the honest one.

Similarly, the scheme is divided into two stages: the distribution stage (see FIG.~\ref{fig:com}(B1)) and the messaging stage (see FIG.~\ref{fig:com}(B2)).

\textbf{Distribution stage}
At this stage, a $2n$-bit string $X_i$ and an $n$-bit string $Y_i$ are generated between the signer and the  receiver $i$ ($i \in \left\{ {1,...,k} \right\}$) through the QKD channel.
At the same time, $k$ $2n$-bit strings $X_{T_1},...,X_{T_k}$ and $k$ $n$-bit strings $Y_{T_1},...,Y_{T_k}$ are generated between the signer and the fixed trusted party by executing QKD.
The signer $U_s$ obtains $k$ $2n$-bit strings
\begin{equation}
	{X_{{T_1}}} \oplus {X_1},...,{X_{{T_k}}} \oplus {X_k}
\end{equation}
and $k$ $n$-bit strings
\begin{equation}
	{Y_{{T_1}}} \oplus {Y_1},...,{Y_{{T_k}}} \oplus {Y_k}
\end{equation}
through the XOR operation.
An authenticated classical channel is required between the fixed trusted party and each receiver.
After this stage, the messaging stage can be executed at any time.

\textbf{Messaging stage}
The signer $U_s$ generates an $n$-bit quantum random number and, together with the key $Y_{T_i} \oplus Y_i$, performs the OTUH process on the $m$-bit message $M$ to generate the digest $D_i$.
Then $U_s$ encrypts the random number and the digest with the $2n$-bit key $X_{T_i} \oplus X_k$ as the signature $S_i$.
Since the signer needs to sign the same message $M$ to $k$ receivers, the above process needs to be executed $k$ times to obtain ($M$, $S_1$), ..., ($M$, $S_k$), sends the $k$ message signature pairs to $k$ receivers respectively.

The receiver $U_i$ receives ($M$, $S_i$) and sends it along with the keys ($X_i$, $Y_i$) to the fixed trusted party over the authenticated classical channel.
After receiving the message from the receiver $U_i$, the fixed trusted party sends the keys ($X_{T_i}$, $Y_{T_i}$) to the receiver $U_i$ over the same authenticated classical channel.
Then the receiver $U_i$ uses the key
$X_{T_i} \oplus X_i$ to decrypt the signature $S_i$ to get the random number and the expected digest $D^i_{exp}$, and uses the key $Y_{T_i} \oplus Y_i$ to perform OTUH process to get the actual digest $D^i_{act}$.
If the two digests are the same, the receiver $i$ declares that the signature ($M$, $S_i$) is accepted, otherwise the signature is rejected.

After the receiver $U_i$ accepts the signature, the fixed trusted party needs to perform the same verification process. Considering that the signature ($M$, $S_i$) may be forged by the receiver $U_i$, the signature received by the fixed trusted party is represented as ($M_i$, $S_{R_i}$).
The fixed trusted party uses ($M_i$, $S_{R_i}$) for OTUH and decryption to get the expected digest $D^{T_i}_{exp}$ and the actual digest $D^{T_i}_{act}$.
If the two abstracts are the same, the fixed trusted party declares the signature accepted, otherwise the signature is rejected.

\textbf{{Key consumption}}
In this scheme, the key required for a signer to sign the message to one receiver is $6n$ bits.
When the number of the receivers is $k$, the total key consumption can be expressed as
\begin{equation}
	{K_{ext}} = 6nk.
\end{equation}

\section{One-time universal hashing}

A hash function $h$ maps a set $A$ to a set $B$, and $\left| A \right| > \left| B \right|$.
If $x, y \in A$ and
$ x \ne y$ have $h(x) = h(y)$, then we say that $x$ and $y$ collide under $h$.
A family H of hash functions $h$ is said to be universal$_2$ if, for every two different $x,y \in A$, its probability of collision $Pr$ under $h$ satisfies
\begin{equation}
	Pr \le \frac{{\rm{1}}}{{\left| B \right|}}.
\end{equation}
Carter and Wegman \cite{CARTER1979143} showed that the family of functions consisting of $n \times m$ Boolean matrices is a universal$_2$ family of hash functions and its collision probability is
\begin{equation}
	Pr=\frac{1}{{{2^n}}}.
\end{equation}
This kind of hash function costs $n \times m$ random bits. In order to reduce the cost, we can restrict the Boolean matrix to be a Toeplitz matrix.
Only $m+n-1$ random bits are needed to construct a Toeplitz matrix, so the use of random bits can obtain a significant savings.
However, it still requires the length of random input bits to be longer than that of the message.

\textbf{LFSR-based Toeplitz hash}
Let $p(x)$ be an irreducible polynomial over GF(2) of degree $n$ and $s_0$, $s_1$, ... be the bit sequence generated by a LFSR with connections corresponding to the coefficients of $p(x)$ and the initial state $s$.
A LFSR-based Toeplitz hash function is defined as the linear combination
\begin{equation}
	h_{p,s}(M)=\oplus _{j = 1}^{m - 1}{M_j} \cdot ({s_j},{s_{j + 1}}...{s_{j + n - 1}})
\end{equation}
for any message
\begin{equation}
	M = M_0 M_1  ... M_{m-1}
\end{equation}
of binary length $m$.
The collision probability of LFSR-based Toeplitz hash \cite{Hugo1994} is
\begin{equation}
	\epsilon=\frac{m}{{{2^{n - 1}}}} .
\end{equation}

\textbf{One-time universal hashing (OTUH)}
This process is performed in the messaging stage of OTUH-QDS.
The signer uses an $n$-bit quantum random number and an $n$-bit keys to generate a LFSR-based Toplitz matrix, and hash the $m$-bit message to generate an $n$-bit digest using the Toplitz matrix.

\section{CW-QKD model}
BBM92-KGP rely on entanglement between distant physical systems, in our case specifically in the polarization degree of freedom of a photon pair.
In an ideal case, entangled photon pairs form a bell state,
e.g.,
\begin{equation}
	\left| {{\phi ^{\rm{ + }}}} \right\rangle {\rm{ = }}\frac{{\rm{1}}}{{\sqrt {\rm{2}} }}({\left| H \right\rangle _A} \otimes {\left| H \right\rangle _B} + {\left| V \right\rangle _A} \otimes {\left| V \right\rangle _B})
\end{equation}
where $H$ ($V$) denotes horizontal (vertical) polarization and the subscripts signify the receiver of the single photon traditionally called Alice ($A$) and Bob ($B$).

\textbf{(1)} \underline{$B$}.
The most general CW-pumped source setup uses a photon source that produces an average number of entangled photon pairs per unit of time (known as brightness \underline{$B$}).
The rate of true coincident counts is given as
\begin{equation}
	C{C^t} = \underline B {\eta _A}{\eta _B}
\end{equation}
where ${\eta _i}$ ($i=A, B$) is overall channel detection probability.

\textbf{(2)} ${e^{pol}}$. We call the probability of erroneous polarization measurement ${e^{pol}}$.
It consists of contributions of the individual polarization error probabilities $e_A^{pol}$ and $e_B^{pol}$ of Alice and Bob, respectively:
\begin{equation}
	{e^{pol}} = e_A^{pol}\left( {1 - e_B^{pol}} \right) + e_B^{pol}\left( {1 - e_A^{pol}} \right).
\end{equation}

\textbf{(3)} $D{C_i}$. $D{C_i}$ is dark count.
The actually measured count rates of Alice and Bob can be written, respectively, as
\begin{equation}
	S_A^m = \underline{B}{\eta _A} + D{C_A},
	S_B^m = \underline{B}{\eta _B} + D{C_B}.
\end{equation}
Here we assume that Alice and Bob each own identical detectors the photon and dark count rates of which can simply be added.

\textbf{(4)} ${t_{CC}}$.
${t_{CC}}$ is coincidence window.
Assuming independent Poissonian photon statistics at Alice and Bob, one can define the mean number of clicks at Alice and Bob, respectively, per coincidence window as
\begin{equation}
	\mu _A^S = S_A^m{t_{CC}},
	\mu _B^S = S_B^m{t_{CC}}.
\end{equation}
The chance of an accidental coincidence being registered can be approximated by the probability of at least one detection event taking place at each of them.
For $\mu _i^S \ll 1$, this probability can be expressed as
\begin{equation}
	{P^{acc}} = (1 - {e^{ - \mu _A^S}})(1 - {e^{ - \mu _B^S}})
	\approx \mu _A^S\mu _B^S.
\end{equation}
where we use the fact that the click probability is given by
$1 - {e^{ - \mu _i^S}}$.
The rate of accidental coincidences per second is therefore
\begin{equation}
	C{C^{acc}} = \frac{{{P^{acc}}}}{{{t_{CC}}}} \approx \frac{{\mu _A^S\mu _B^S}}{{{t_{CC}}}} = S_A^mS_B^m{t_{CC}}.
\end{equation}

\textbf{(5)} $\eta ^{{t_{CC}}}$.
The proportion of true coincidences which fall into the chosen coincidence window $t_{CC}$ can be expressed as the integration
\begin{equation}
	{\eta ^{{t_{CC}}}} = \int_{ - {t_{CC}}/2}^{{t_{CC}}/2} {j(t,{t_\Delta },{t_D} = 0)dt = } erf[\sqrt {\ln (2)} \frac{{{t_{CC}}}}{{{t_\Delta }}}]
\end{equation}
where $t_\Delta$ is the resulting timing imprecision between Alice's and Bob's measurements, $t_D$ is a certain constant delay and
$j(t,{t_\Delta },{t_D})$ is a normal distribution
\begin{equation}
	j(t,{t_\Delta },{t_D}) = \frac{2}{{{t_\Delta }}}\sqrt {\frac{{\ln (2)}}{\pi }} \exp \left[ { - \frac{{4\ln (2)}}{{t_\Delta ^2}}{{\left( {t - {t_D}} \right)}^2}} \right]
\end{equation}
of the $g^{(2)}$ intensity correlation with the full width at half maximum $t_\Delta$ between Alice's and Bob's detectors.
The actually measured coincidences can be defined as
\begin{equation}
	C{C^m} = {\eta ^{{t_{CC}}}}C{C^t} + C{C^{acc}}.
\end{equation}
This is the total number of detector events per second that Alice and Bob use to create their key.
The subset of these events occurs at the rate
\begin{equation}
	C{C^{err}} = {\eta ^{{t_{CC}}}}C{C^t}{e^{pol}} + \frac{1}{2}C{C^{acc}}.
\end{equation}

\textbf{(6)} Error rate and secure key rate.
The quantum bit error rate can be calculated as
\begin{equation}
	E = \frac{{C{C^{err}}}}{{C{C^m}}} = \frac{{{\eta ^{{t_{CC}}}}C{C^t}{e^{pol}} + \frac{1}{2}C{C^{acc}}}}{{{\eta ^{{t_{CC}}}}C{C^t} + C{C^{acc}}}}.
\end{equation}
The final amount of achievable key per second can be evaluated as
\begin{equation}
	{R^S} = qC{C^m}\left[ {1 - f\left( {{E_{bit}}} \right){H_2}\left( {{E_{bit}}} \right) - {H_2}\left( {{E_{ph}}} \right)} \right],
\end{equation}
where $H_2$ is the binary entropy function defined as
\begin{equation}
	{H_2}\left( x \right) =  - x{\log _2}\left( x \right) - \left( {1 - x} \right){\log _2}\left( {1 - x} \right).
\end{equation}

\end{document}